\documentclass[11pt]{article}      
\author{Kris Krogh \\
Neuroscience Research Institute\\ University of California, Santa
Barbara, CA 93106, USA\\ email: k\_krogh@lifesci.ucsb.edu}

\title
{Equations of Motion in a Quantum-mechanical Theory of Gravity}

\date{August 11, 2005}

\begin{document}

\maketitle

\begin{abstract}

\normalsize  An earlier paper~\cite{kk1} presented a gravity theory
based on the optics of de Broglie waves rather than curved
space-time. While the universe's geometry is flat, it agrees with
the standard tests of general relativity.  A second paper~\cite{kk2}
showed that, unlike general relativity, it agrees with Doppler
tracking signals from the Pioneer 10 and 11 space probes. There a
gravitational acceleration equation plays an important role,
accounting for the relative motions of Earth and the probes. Here
it's shown that equation also describes Mercury's orbit.

\end{abstract}

\vfill\eject

\section{Introduction}

A previous paper~\cite{kk1} introduced a theory of gravity based on
the optics of de Broglie waves, without curved space-time.  While
agreeing with the standard experimental tests of general relativity,
the theory is inherently quantum-mechanical.  A second
paper~\cite{kk2} showed that, unlike general relativity, it agrees
with the observed motions~\cite{alllnt} of the Pioneer 10 and 11
space probes.

Many-body systems are described by a gravitational acceleration
equation derived in the first paper.  That was used to calculate a
perturbation of the Moon's orbit observed in the lunar ranging
experiment.  And subsequently to obtain the Pioneer motions. It also
gives Mercury's observed orbit.  Since the acceleration equation is
a pivotal one, we'll demonstrate that here.

Like the quantum mechanics (``wave mechanics") of de Broglie and
Schr\"{o}din-ger, the development of this gravity theory was guided
by the optical-mechanical analogy~\cite{tlh}.  In optics, to fully
describe the propagation of light waves and photons, it's necessary
to calculate wave amplitudes, as given by Huygens' principle. In the
short-wavelength limit, where diffraction and interference effects
can be ignored, geometrical optics can be used.

De Broglie~\cite{ldb} and Schr\"{o}dinger~\cite{es} found the same
holds for matter and matter waves.  Where the de Broglie wavelengths
of particles are sufficiently short, quantum mechanics reduces to
ordinary Hamiltonian mechanics.  As Hamilton discovered earlier, the
latter is directly analogous to geometrical optics.

Rather than curving space-time, the effect of gravitational
potentials in this theory is a slowing of quantum mechanical waves
-- both matter and light.  (Where Einstein assumed a constant speed
of light and variable space and time, the assumption here is the
opposite.) From de Broglie, the velocity $V$ of matter waves is
\begin{equation}
V \:=\: \frac{\,c^2}{\,v}
\end{equation}
where $c$ is the speed of light, and $v$ is the velocity of an
associated particle or body.

Here the speeds of light and de Broglie waves in a gravitational
potential $\Phi_g$ are given by
\begin{equation}
\frac{\,c}{\,c_0} \:=\: \frac{\,V}{\,V_0} \:=\; e^{2 \Phi_{\!g} /
c_0^2}
\end{equation}
where the subscript $0$ represents the same quantity in the absence
a gravitational potential.  The reduced wave velocities are
manifested equally in their wavelengths and frequencies, as
\begin{equation}
\frac{\,\nu}{\,\nu_0} \;=\: \frac{\,\lambda}{\,\lambda_0} \;=\:
e^{\Phi_{\!g} / c_0^2}
\end{equation}

Since the frequencies of de Broglie waves in atoms determine the
rates of clocks, and their wavelengths the sizes measuring rods,
both diminish in gravitational potentials.  Consequently, in accord
with Poincar\'{e}'s principle of relativity, the reduced wave
velocities aren't apparent to a local observer.  Time and space
themselves are unaffected, and the latter remains isotropic, in
agreement with the observed large-scale flatness of the universe.

It's pointed out, for example by Wheeler~\cite{wf}, that not only
elementary particles, but atoms and macroscopic objects have de
Broglie wavelengths.  An atomic electron is subject to the condition
that its orbit is an integral number of those wavelengths. In this
theory, the same condition applies in principle to orbits of
astronomical bodies.

Of course the $h/p$ de Broglie wavelength for an orbiting body is
many orders of magnitude shorter than an electron's. And since an
orbit's size is also much larger, its quantization is unobservable.
It follows the short-wavelength approximation can be used to
describe astronomical orbits, by the methods of geometrical optics.
(Determining trajectories orthogonal to the de Broglie wavefronts.)
That was found to give an orbit equation which agrees with
Mercury's~\cite{kk1}.

The relativistic Lagrangian for a charged particle in
electromagnetic potentials is
\begin{equation}
L \:=\, \,-\,m_0 c^2{\sqrt{1\,-\,v^2\!/c^2}} \:- \:q\,(\,\Phi -\,
{\bf v} \cdot {\bf A}\,/\,c\,) \label{eq:4}
\end{equation}
where $m_0$ is its rest mass, $q$ the charge, and $\Phi$ and $A$ are
the scalar and vector potentials.  This is related directly to its
de Broglie frequency~\cite{kk1}.  In this theory, the Lagrangian for
a particle or body in a gravitational potential becomes
\begin{equation}
L \,= \left(-\,m_{00} c_0^2\,{\sqrt{1\,-\,v^2\!/c^2}} \:-
\:q_0\,(\,\Phi_0 -\, {\bf v} \cdot {\bf A}_0\,/\,c\,)\right)
e^{\Phi_{\!g}/c_0^2} \label{eq:5}
\end{equation}
with $m_{00}$, $q_0$, $\Phi_0$ and $A_0$ the corresponding
quantities in the absence of a gravitational potential.

The resulting Euler-Lagrange equation of motion for a body in a
central gravitational field gives the same first-order differential
equation of the orbit found by de Broglie wave optics~\cite{kk1}.
From that, this second-order equation was derived
\begin{equation}
\frac {\,d^2 r} {\,d \theta ^2} \:=\: \frac {\,2 - \mu/r} {r}
\left(\frac {\,dr} {\,d \theta} \right)^{\!2} \!+\, r \,-\, \mu
\,-\, \frac {\,r^2 \mu \,e^{4\mu/r}} {k^2} \label{eq:6}
\end{equation}
where $r$ and $\theta$ are polar coordinates.  Here $\mu$ is given
by
\begin{equation}
\mu \:=\: GM \, / \,c_0^2 \label{eq:7}
\end{equation}
with $G$ the gravitational constant and $M$ the mass of the central
body.  And $k$ is a constant of the orbit
\begin{equation}
k \:=\: \,r^2 \,\dot \theta \:e^{4\mu/r} / \,c_0 \label{eq:8}
\end{equation}
with the dot indicating time differentiation.

For comparison, in this notation the corresponding Newtonian
equation is
\begin{equation}
\frac {\,d^2 r} {\,d \theta ^2} \:=\: \frac {\,2} {\,r} \left(\frac
{\,dr} {\,d \theta} \right)^{\!2} \!+\, r \,-\, \frac {\,r^2 \mu }
{\,k^2}
\end{equation}
where $k$ again is a constant of the orbit
\begin{equation}
k \:=\: r^2 \,\dot \theta \,/ \,c_0
\end{equation}
representing the body's conserved angular momentum divided by $c_0$.

The Lagrangian of Eq.~(\ref{eq:4}) gives the Lorentz equation for
the electromagnetic force $\bf{F}$ on a charged particle
\begin{equation} {\bf F}\:=\: q \left( {\bf E} + \frac{{\bf
v} \times {\bf B}}{c} \, \right) \label{eq:11}
\end{equation}
where the electromagnetic potentials are translated into the
electric and magnetic fields ${\bf E}$ and ${\bf B}$. This equation
can be used to determine the motions of arbitrary systems of
charges, when their energies aren't conserved.  No comparable
equation exists for gravity in general relativity. Instead, an
approximate many-body Lagrangian is used to describe many-body
systems~\cite{kn}.

In this non-metric theory, the Lagrangian in Eq.~(\ref{eq:5}) {\em
does} give a gravitational counterpart of the Lorentz force
equation~\cite{kk1}
\begin{equation}
{\bf a} \:=\: -\nabla\Phi_{\!g} \left(e^{4\Phi_{\!g}/c_0^2}+ \frac
{\,v^2}{\,c_0^2}\right) + \frac{\,4{\bf v}}{\,c_0^2} \left(\frac
{\,d{\Phi_{\!g}}}{\,dt}\right) \label{eq:12}
\end{equation}
where ${\bf a}$ represents a body's acceleration.  (Multiplying both
sides by its mass would give a gravitational force.)  Like
Eq.~(\ref{eq:11}), this can be used iteratively to calculate the
motions of arbitrary systems of bodies whose energies aren't
conserved.  Below we'll derive an orbit equation from this equation
of motion, and compare it to that already obtained from de Broglie
wave optics and the Euler-Lagrange equations of motion.

\section{Orbits from the Acceleration Equation}

For a body orbiting in a central gravitational field, from
Eq.~(\ref{eq:12}), its radial acceleration can be expressed in
isotropic polar coordinates as
\begin{equation}
\ddot{r} \,-\, r \dot{\theta}^2 \:=\:  -\nabla\Phi_{\!g}
\left(e^{4\Phi_{\!g}/c_0^2}+ \frac {\,r^2 \dot{\theta}^2+
\,\dot{r}^2}{\,c_0^2}\right) + \frac{\,4{\dot{r}}}{\,c_0^2}
\left(\dot{r} \,\nabla\Phi_{\!g}\right) \label{eq:13}
\end{equation}
(See Becker~\cite{rab}.)

In this theory, the gravitational potential due to a stationary
spherical body is just
\begin{equation}
\Phi_{\!g} \:=\: -\,GM / r
\end{equation}
From the notation of  Eq.~(\ref{eq:7}),
\begin{equation}
\Phi_{\!g}/c_0^2 \:=\:  -\mu/r
\end{equation}
and the gradient of the gravitational potential is
\begin{equation}
\nabla\Phi_{\!g} \:=\:  c_0^2\,\mu/r^2
\end{equation}
After these substitutions, Eq.~(\ref{eq:13}) becomes
\begin{equation}
\ddot{r} \,-\, r \dot{\theta}^2 \:=\:  -\frac{\, c_0^2 \mu}{\,r^2}
\,e^{-4 \mu/r} - \,\mu \dot{\theta}^2 \,+ \frac{\,3\mu
\dot{r}^2}{\,r^2} \label{eq:17}
\end{equation}

The orbiting body's radial velocity can be expressed in terms of
$\dot \theta$ as
\begin{equation}
\dot r \:=\: \frac {\,d r}{\,d \theta} \,\dot\theta\ \label{eq:18}
\end{equation}
Taking the time derivative,
\begin{eqnarray}
\ddot r & = & \frac{\,d}{\,dt} \left( \! \frac{\,dr}{\,d \theta} \,\dot\theta \right ) \nonumber\\
 & = & \frac{\,d^2 r}{\,d \theta^2} \,\dot\theta^2 \,+\, \frac{d r}{d
 \theta} \frac{\,d \dot\theta}{\,d r} \,\dot r \label{eq:19}
\end{eqnarray}

Again, this theory gives a conserved quantity $k$ for an orbiting
body, similar to angular momentum.  Rearranging Eq.~(\ref{eq:8}) for
$k$ gives
\begin{equation}
\dot \theta \:=\: \frac {\,k \,c_0 \,e^{-4\mu/r}} {\,r^2}
\label{eq:20}
\end{equation}
Differentiating that with respect to $r$, then putting the result in
terms of $\dot\theta$,
\begin{equation}
\frac{\,d \dot \theta}{\,d r} \:=\: - \!\left(
\frac{\,2}{\,r}\,-\frac{\,4 \mu}{\,r^2}\right) \dot\theta
\end{equation}
Substituting for $d\dot\theta/dr$ and $\dot r$ in Eq.~(\ref{eq:19}),
we get
\begin{equation} \ddot r \:=\: \left(\frac{\,d^2r}{\,d\theta^2}\,-
\left( \frac{\,2}{\,r}\,-\frac{\,4 \mu}{\,r^2}\right) \!\left(\frac
{\,d r}{\,d \theta} \right)^{\!2} \right) \dot\theta^2
\end{equation}

After substituting for $\dot r ^2$ in Eq.~(\ref{eq:17}), we also get
\begin{equation}
\ddot{r} \:=\:  -\frac{\, c_0^2 \mu  e^{-4 \mu/r}}{\,r^2}  \,+\, r
\dot\theta^2 \,- \,\mu \dot{\theta}^2 \,+ \frac{\,3\mu }{\,r^2}
\left(\!\frac{\,d r}{\,d\theta} \right)^{\!2} \dot\theta^2
\end{equation}
Equating the right sides of the last two equations, dividing by
$\dot\theta^2$, and rearranging gives
\begin{equation}
\frac{\,d^2r}{\,d\theta^2} \:=\: \frac{\,2-\mu/r}{\,r}
\left(\!\frac{\,d^2r}{\,d\theta} \right)^{\!2} + \,r \,-\, \mu
-\frac{\,c_0^2 \mu e^{-4 \mu/r}}{\,r^2 \,\dot\theta^2}
\end{equation}
Finally, substituting for $\dot \theta ^2$ from  Eq.~(\ref{eq:20})
gives the second-order differential equation of the orbit found
previously, Eq.~(\ref{eq:6}).

The initial paper showed Eq.~(\ref{eq:6}) is the derivative of this
one
\begin{equation}
\frac {\,dr} {\,d \theta} \:=\: \frac {r \sqrt{\,r^2 \left(
e^{4\mu/r} - \frac {E_{00}^2} {E^2} \,e^{2\mu/r} \right) -k^2}} {k}
\label{eq:127}
\end{equation}
where $E$ and $E_{00}$ are constants of the orbit, corresponding to
the body's total relativistic energy and its rest energy in the
absence of a gravitational potential.  This first-order equation was
derived from both de Broglie wave optics and the Euler-Lagrange
equations of motion.  And it was shown to agree with Mercury's
orbit~\cite{kk1}.

\section{Conclusions}

Mercury's orbit remains one of the strongest tests of general
relativity. The same orbital precession is predicted by this
quantum-mechanical theory.  We've arrived at a single orbit equation
by three routes now: Directly from de Broglie wave optics, the
Euler-Lagrange equations of motion, and Eq.~(\ref{eq:12}) for
gravitational acceleration.

Unlike the Euler-Lagrange equations, the acceleration equation
describes the motion of bodies whose energies aren't conserved. It
also describes the relative motions of Earth and the Pioneer 10 and
11 space probes~\cite{kk2}, which are unexplained by general
relativity.

\section*{Acknowledgments}

I thank Kenneth Nordtvedt for originally raising the issue of a
many-body Lagrangian for this gravity theory.  And Stan Robertson
and Martin Smith for helpful discussions.

\vfill\eject

\end{document}